\documentclass[letterpape,final]{IEEEtran}
\usepackage{booktabs}
\usepackage{amsmath}
\usepackage{amsthm}
\usepackage{mathrsfs}
\usepackage{mathtools}
\usepackage[final,markup=underlined,defaultcolor=red,authormarkuptext=name,authormarkup=none]{changes}
\definechangesauthor[name={Lisheng Fan}, color=red]{A1} %修订作者1
\definechangesauthor[name={Shunpu Tang}, color=blue]{A2} %修订作者2
\usepackage{caption}
\usepackage{subcaption}
\usepackage{graphicx}
\usepackage{latexsym}
\usepackage{amssymb,amsbsy,verbatim,array}
\usepackage{epsfig}
\usepackage{epstopdf}
\usepackage{array}

\usepackage{blindtext}
\usepackage{cite} 
\usepackage{color}
\usepackage[ruled,lined,linesnumbered]{algorithm2e}
\usepackage{makecell}
\usepackage{extarrows}
\usepackage{multicol}
\usepackage{multirow}
\usepackage{bm}
\usepackage{amsthm}
\usepackage{mathrsfs}
\usepackage{booktabs}
\usepackage{siunitx}
\usepackage[switch]{lineno}
\usepackage{tabularx,booktabs}
\usepackage[flushleft]{threeparttable}
\usepackage[flushleft]{threeparttable}
\usepackage[short]{optidef}

\usepackage[colorlinks,
            linkcolor=black,      
            anchorcolor=blue,  
            citecolor=green,     
            ]{hyperref}

\begin{document}

\title{Contrastive Learning based Semantic Communication for Wireless Image Transmission}
\author{
    Shunpu Tang, 
    Qianqian Yang, 
    Lisheng Fan, 
    Xianfu Lei,
    Yansha Deng,
    and Arumugam Nallanathan, \IEEEmembership{Fellow, IEEE} 
    \vspace{-5mm}
    
    \thanks{S. Tang and L. Fan are both with the School of Computer Science and Cyber Engineering, Guangzhou University, Guangzhou, China (e-mail: tangshunpu@e.gzhu.edu.cn, lsfan@gzhu.edu.cn). }
    \thanks{Q. Yang is with the College of Information Science and Electronic Engineering, Zhejiang University, Hangzhou, China (e-mail: qianqianyang20@zju.edu.cn).}
    \thanks{X. Lei is with the Institute of Mobile Communications, Southwest Jiaotong University, Chengdu, China (e-mail: xflei@home.swjtu.edu.cn).}
    \thanks{Y. Deng is with the Department of Engineering, Kings College London, London, U.K (e-mail:yansha.deng@kcl.ac.uk)}

    % \thanks{W. Xu is with the National Mobile Communications Research Laboratory, Southeast University, Nanjing 210096, China. (e-mail: wxu@seu.edu.cn)}
    \thanks{A. Nallanathan is with the School of Electronic Engineering and Computer Science, Queen Mary University of London, London, U.K (e-mail: a.nallanathan@qmul.ac.uk).}
}
\maketitle

\begin{abstract}
Recently, semantic communication has been widely applied in wireless image transmission systems as it can prioritize the preservation of meaningful semantic information in images over the accuracy of transmitted symbols, leading to improved communication efficiency. However, existing semantic communication approaches still face limitations in achieving considerable inference performance in downstream AI tasks like image recognition, or balancing the inference performance with the quality of the reconstructed image at the receiver. Therefore, this paper proposes a contrastive learning (CL)-based semantic communication approach to overcome these limitations. Specifically, we regard the image corruption during transmission as a form of data augmentation in CL and leverage CL to reduce the semantic distance between the original and the corrupted reconstruction while maintaining the semantic distance among irrelevant images for better discrimination in downstream tasks. Moreover, we design a two-stage training procedure and the corresponding loss functions for jointly optimizing the semantic encoder and decoder to achieve a good trade-off between the performance of image recognition in the downstream task and reconstructed quality. Simulations are finally conducted to demonstrate the superiority of the proposed method over the competitive approaches. In particular, the proposed method can achieve up to 56\% accuracy gain on the CIFAR10 dataset when the bandwidth compression ratio is 1/48.

\end{abstract}

\section{Introduction}
Recently, semantic communication has emerged as a promising approach for efficient  image transmission in wireless network, and attracted increasing research interests from academic. Compared with the conventional communication paradigm based on Shannon's theory, semantic communication aims to prioritize to preserve meaningful semantic information  over focusing on the accuracy of transmitted symbols, which can significantly reduce the amount of data to be transmitted and improve the communication efficiency\cite{Semantic_survey}. 

A major challenge in semantic communication for image transmission is to effectively extract semantic information at the transmitter while accurately reconstructing it at the receiver under limited communication conditions. To overcome this issue, deep learning (DL) has been applied in semantic communication, due to its growing success in semantics extraction and reconstruction. In this direction, the authors in \cite{DeepJSCC,jscc_f} proposed a DL base joint source-channel coding (DeepJSCC), where the encoder and decoder were designed based on autoencoder and jointly optimized for semantic information transmission to achieve a good image reconstruction quality. Moreover,  a collaborative training framework for semantic communication was proposed in \cite{DeepSC}, where users could train their semantic encoder to improve the performance of unknown downstream inference tasks. However, these approaches still suffer from limitations in achieving considerable inference performance and making a good trade-off between it and image reconstruction quality according to communication conditions. 

In essence, the semantic distance between two irrelevant images is large due to different main semantic information, while the distance is small enough between two nearly identical images sharing the same semantic information. Motivated by this fact, we propose to integrate contrastive learning (CL) with semantic communication and design a two-stage training procedure. The key design in this framework is that we regard the image corruption that occurs during transmission over a limited channel as a form of data augmentation operation in \cite{SIMLR} and propose the semantic contrastive coding to reduce the semantic distance (a.k.a semantic similarity) between the original and reconstructed image while maintain the semantic distance among irrelevant images for better discrimination. Based on contrastive loss, we design  the two-stage training procedure and loss functions for jointly optimizing the encoder and decoder, thereby achieving a good balance between the inference performance in the downstream task and reconstruction quality. Simulations are finally conducted to demonstrate the superiority of the proposed method over the competitive approaches. 

%  The main semantic information between two irrelevant images is greatly different, which results in a substantial semantic distance between them, while the semantic distance  between a pair of nearly identical images is small enough, as they share the same semantic information. Motivated by this fact, we propose to integrate contrastive learning (CL) with semantic communication and design a two-stage training procedure to address the limitations of the existing approaches. The key design in this framework is that we regard the image corruption that occurs during transmission over a limited channel as a form of data augmentation operation in \cite{SIMLR} and utilize the CL to reduce the semantic distance (a.k.a semantic similarity) between the original and reconstructed image while maintain the semantic distance among irrelevant images for better discrimination. Based on contrastive loss, we design  two-stage training procedure and loss functions for jointly optimizing the encoder and decoder, thereby achieving a good balance between the downstream task and reconstruction quality. Simulations are finally conducted to demonstrate the superiority of the proposed method over the competitive approaches. 

\section{System model}
This paper investigates a semantic communication system for wireless image transmission, where a convoloutional neural network (CNN) based semantic encoder and decoder \replaced{are}{is} deployed in the transmitter and receiver, respectively. The semantic encoder is used to extract the semantic information of input image  $\bm{x} \in \mathbb{R}^{c \times h\times w}$ and directly realize the non-linear mapping from semantic information into the $k$-dim complex-valued vector $\tilde{\bm{s}} \in \mathbb{C}^{k}$, given by
\begin{equation}
    \tilde{\bm{s}}=\mathcal{E}_{\theta_1}(\bm{x}),
\end{equation}
where $\mathcal{E}_{\theta_1}(\cdot)$ represents the semantic encoding operation with parameter $\theta_1$, and $c$, $h$, and $w$ denote the number of channels, \deleted{the} height, and \deleted{the} width of the image, respectively. To simplify, we use $n=c \times h\times w$ to stand for the dimension of $\bm{x}$. Typically, $k<n$ should be satisfied to meet the bandwidth constraint, and  $k/n$ is referred to as the bandwidth compression ratio. \added{In particular,} \replaced{a}{A} large bandwidth compression ratio indicates a favorable communication condition, whereas a small one denotes a limited usage of bandwidth. In addition,  a power normalization layer\cite{DeepJSCC} is used at the end of the semantic encoding network to satisfy the average power constraint $\frac{1}{k}\mathbb{E}[\bm{s}^* \bm{s}]\leq P$ at the transmitter, which can be written as
 \begin{equation}
    \bm{s}=\sqrt{kP}\frac{\tilde{\bm{s}}}{\sqrt{\tilde{\bm{s}}^* \tilde{\bm{s}}}},
\end{equation}
where $\bm{s}$ is the channel input signal that meets the power constraint, and \replaced{$^{*}$}{${*}$} denotes the conjugate transpose. Next, $\bm{s}$ will be transmitted over the additive white Gaussian (AWGN) channel, given by
\begin{equation}
    \hat{\bm{s}}=\bm{s}+\bm{\epsilon},
\end{equation}
where $\hat{\bm{s}}$ is the received signal, and $\bm{\epsilon} \in \mathbb{C}^{k}$ denotes the independent and identically distributed (IID) channel noise sample, which follows symmetric complex Gaussian distribution $\mathcal{CN}(0,\sigma^2\bm{I})$ with zero mean and variance $\sigma^2$.

%  After that, the reconstructed image will be  sent to the downstream task's backbone and classifier for further processing and obtaining the final inference results.  

% Specifically, the input image at the transmitter can be denoted as $\bm{x} \in \mathbb{R}^{c \times h\times w}$, where $c=3$ for an RGB image in the practical system, and $h$ and $w$ represents the height and width of the image, respectively. To simplify, we use $n=c \times h\times w$ to stand for the original input dimension. At the transmitter, a semantic coding network is deployed to extract the important semantic information, 
% and 

\begin{figure}[t!]
    \centering
    \includegraphics[width=3.4in]{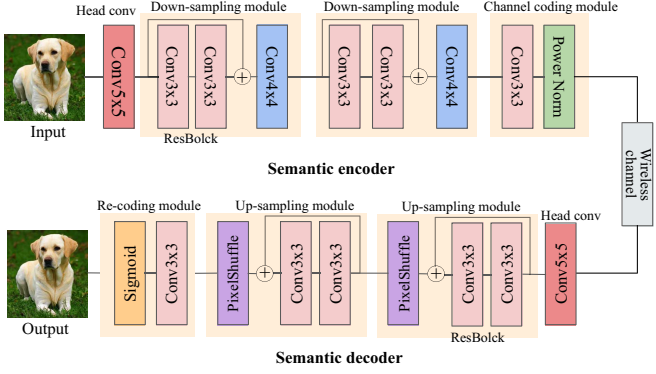}
    \caption{Network architecture of the semantic encoder and decoder.}
    \label{fig:network}
\end{figure}

The semantic decoder deployed at the receiver will reconstruct the original image $\hat{\bm{x}}\in \mathbb{R}^{c \times h\times w}$ from  $\hat{\bm{s}}$ according to
\begin{equation}
    \hat{\bm{x}}=\mathcal{D}_{\theta_2}(\bm{\hat{s}}),
\end{equation}
where $\mathcal{D}_{\theta_2}(\cdot)$ is the semantic decoding operation parameterized by $\theta_2$. Subsequently, $\hat{\bm{x}}$ will be used to exert downstream task and obtain the inference results through the following process
% Subsequently, $\hat{\bm{x}}$ will be fed into the backbone $\mathcal{F}^{b}_{\phi_1}$ with parameter ${\phi_1}$ and classifier $\mathcal{F}^{cls}_{{\phi_2}}$ paramized by $\phi_2$ of the downstream task for further inference, through the following process
\begin{equation}
\label{eq::backbone}
    \bm{f}_{\bm{x}}=\mathcal{F}^{b}_{\phi_1}(\hat{\bm{x}}),
\end{equation}
where $\mathcal{F}^{b}_{\phi_1}(\cdot)$ characterized by parameter ${\phi_1}$ denotes the feature extraction operation performed by the CNN backbone of downstream task, and $\bm{f}_{\bm{x}}=\{\bm{f}^{(1)},\bm{f}^{(2)},\cdots \bm{f}^{(C)}\}$ is the output feature map with $C$ channels. The inference results $\hat{\bm{y}}$ can be obtained by passed $\bm{f}_{\bm{x}}$ to the classifier  $\mathcal{F}^{cls}_{{\phi_2}}(\cdot)$ with parameter $\phi_2$, which can be expressed as
\begin{equation}
    \hat{\bm{y}}=\mathcal{F}_{\phi_2}^{cls}(\bm{f}_x).
\end{equation}
Since maintaining semantic information in the reconstructed image is crucial for the inference performance, especially when \added{the} channel bandwidth is limited. Therefore,  \added{it is of vital importance to design} the semantic encoder and decoder, as well as the training procedure\replaced{.}{ require careful design.}

\section{Proposed Framework}
\begin{figure}[t!]
    \centering
    \includegraphics[width=3.2in]{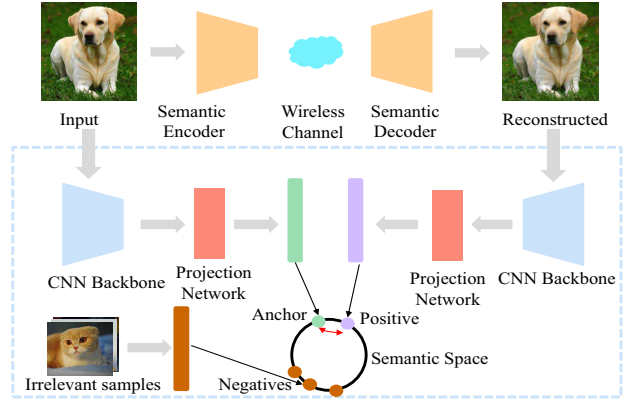}
    \caption{Illustration of the proposed semantic contrastive coding.}
    \label{fig:contrastive}
\end{figure}
In this section, we will introduce the proposed CL based semantic communication framework. Specifically, we first present the architecture of the semantic encoder and decoder, and then we will provide the details of semantic contrastive coding and the training procedure.
\subsection{Architecture of Semantic Encoder and Decoder}

% \begin{figure*}[ht!]
%     \centering
%     \includegraphics[width=5.5in]{figs/CL.pdf}
%     \caption{Caption}
%     \label{fig:contrastive}
% \end{figure*}

The network architecture of the semantic encoder and decoder plays a critical role in the extraction of semantic information. Therefore,  we do not utilize the straightforward approach of stacked convolutional layers in \cite{DeepJSCC}, as this simple architecture lacks this ability. The \deleted{used} architecture of the \added{proposed} semantic encoder and decoder are presented in \autoref{fig:network}. The semantic encoder comprises a $5\times 5$  head convolution, two down-sampling modules, and a channel coding module. Each down-sampling module includes a basic block in ResNet\cite{ResNet} (we refer to as ResBolck) for capturing the spatial feature\deleted{s} of the image, and a $4\times 4$ convolution with stride $2$ for down-sampling the image. The channel coding module is used to mitigate channel corruption and output the $k$-dim complex-valued channel input that satisfies the bandwidth and power constraint. 

Moreover, we adopt a symmetrical architecture in the decoder, which consists of a $5\times 5$ head convolution, two up-sampling modules, and a re-coding module. In the up-sampling module, ResBolcks are also used as in the encoder and we adopt the Pixel-Shuffle technology\cite{Pixelshuffle} to up-sample the image, as it can provide a more efficient computing paradigm and better reconstruction performance compared to transposed convolution used in \cite{DeepJSCC}.  The re-coding module consists of a $3\times 3$ convolution followed by the Sigmoid activated function to generate the reconstructed image. Notably, \added{the} batch normalization and \deleted{the} parametric rectified linear unit (PReLU) activated function are followed with all convolutions, \replaced{if not specified}{unless otherwise noted}.
\subsection{Semantic Contrastive Coding}

The key design of semantic contrastive coding is  inspired by the success of CL  which employs data augmentation to generate samples with similar vision representation and minimizes the distance among them to pretrain the backbone. We modify the CL process to adapt it for the semantic communication system.  Specifically, we replace the data augmentation with the process of wireless transmission, as the image corruption that occurs during the transmission can be viewed as a form \added[id=A2]{of} data augmentation, and the original image and reconstructed one should keep a small semantic distance for \replaced{an}{a} efficient semantic communication system. Moreover, we utilize a pretrained backbone to extract features and a learnable projection network to map these features into the semantic space. In further, by incorporating the contrastive loss in semantic space, we jointly optimize the semantic encoder and decoder rather than pretraining the backbone in CL.

The details of the proposed semantic contrastive coding \replaced{are}{is} shown in Fig. \ref{fig:contrastive}. The process begins with the semantic encoding and decoding for a typical image $\bm{x}$ in a training batch $\mathcal{B}$, where we can obtain the reconstructed $\hat{\bm{x}}$.  The backbone network $\mathcal{F}^{b}_{\phi_1}(\cdot)$ is applied to $\bm{x}$ and $\hat{\bm{x}}$, which generates the feature maps $\bm{f}_{\bm{x}}=\mathcal{F}^{b}_{\phi_1}(\bm{x})$ and $\bm{f}_{\hat{\bm{x}}}=\mathcal{F}^{b}_{\phi_1}(\hat{\bm{x}})$, respectively.
% After that, we introduce a projection head $\mathcal{P}(\cdot)$ to map the feature map into semantic space, which is realized by a learnable multi-layer perception. 
Next, a fully connected projection network $\mathcal{P}_{\psi}(\cdot)$ with learnable parameter $\psi$ followed by a normalization operation maps the features into a semantic space defined as a hypersphere. During the training stage, $\mathcal{P}_{\psi}(\cdot)$ can be updated to enhance the understanding of features, thereby learning the mapping from features to semantics. Specifically, the projected results of $\bm{f}_{\bm{x}}$ and $\bm{f}_{\hat{\bm{x}}}$ can be represented as $\bm{q}_{\bm{x}}=\mathcal{P}_{\psi}({\bm{f}_{\bm{x}}})$ and $\bm{v}_{+}=\mathcal{P}_{\psi}({\bm{f}_{\hat{\bm{x}}}})$, respectively, where $\bm{q}_{\bm{x}}$ is referred to as the \textit{anchor}, and $\bm{v}_{+}$ is called the \textit{positive}. We can use the cosine similarity  between \textit{anchor} and \textit{positive} to define the semantic distance between $\bm{x}$ and $\hat{\bm{x}}$.

% In other words, the semantic distance is modeled as the distance between $\bm{q}_{\bm{x}}$ and $\bm{v}_{+}$ on the hypersphere.

For the remaining samples $\bm{m} \in \mathcal{B}/\{\bm{x}\}$ within training batch $\mathcal{B}$, the same procedures will be followed. Specifically, we can obtain the feature map $\bm{f}_{\bm{m}} =\mathcal{F}_b (\bm{m})$ by feding  $\bm{m}$ into the backbone network, and then project $\bm{f}_{\bm{m}}$ into the semantic space using $\bm{v}_{\bm{m}}=\mathcal{P}_{{\psi}}(\bm{f}_m )$, where we refer $\bm{v}_{\bm{m}}$ to as the \textit{negative}. Similar, the semantic distance between $\bm{x}$ and $\bm{m}$ can be defined as the cosine similarity  between \textit{anchor} and \textit{
negative}. The objective of semantic contrastive coding is to minimize the semantic distance between the original and reconstructed image while maximizing the semantic distance among the original image and the irreverent images. Therefore, we can use the InfoNCE function\cite{SIMLR} to define the semantic contrastive loss for the training batch $\mathcal{B}$, which can be expressed as
% \begin{equation}
%     \mathcal{L}_{semantic}=\mathbb{E}_{x\in \mathcal{B} } \bigg \{- \log frac{\exp ( \text{SOFTMAX}(<\bm{q}_{\bm{x}},\bm{v}_{+}>/\tau))}{\sum_{\bm{m} \in \mathcal{B}/\{\bm{x}\}}  \exp(\text{SOFTMAX}(<\bm{q}_{\bm{x}},\bm{v}_m>/\tau))} \bigg \}
% \end{equation}
\begin{equation}
    \mathcal{L}_{sem}=\mathbb{E}_{\bm{x} \in \mathcal{B} } \bigg \{- \log \frac{\exp (\bm{q}_{\bm{x}}\cdot\bm{v}_{+}/\tau)}{\sum_{\bm{m} \in \mathcal{B}/\{\bm{x}\}}  \exp(\bm{q}_{\bm{x}}\cdot\bm{v}_m/\tau)} \bigg \},
\end{equation}
where $\tau\added{>0}$ is the temperature coefficient used to smooth the probability distribution. Next, we will introduce how to take into account the semantic contrastive coding and semantic contrastive loss to design the loss function and training procedure.

\subsection{Loss Function and Training Procedure}
Based on the semantic contrastive coding, we design a two\added{-}stage training strategy to optimize the semantic encoder and decoder. The first stage is pre-training, where we employ the semantic contrastive coding approach to train the weights of encoder $\theta_1$, decoder $\theta_2$ and project network $\psi$ simultaneously. However, it is difficult to achieve a fast convergence speed when we only optimize semantic contrastive loss. To tackle this issue, we combine it with the reconstructed loss between $\bm{x}$ and $\hat{\bm{x}}$, since \replaced{reducing}{reduce} the reconstructed loss can help improve the convergence speed in the early training rounds. Specifically, we apply the mean square error (MSE) function to evaluate the reconstruction loss for training batch $\mathcal{B}$, which can be expressed as
\begin{equation}
\mathcal{L}_{rec}=\mathbb{E}_{\bm{x} \in \mathcal{B}} \bigg \{\frac{1}{n}||\bm{x}-\hat{\bm{x}}||^2_2 \bigg \}.
\end{equation}
Therefore, the loss function in the first training stage can be summarized as the linear combination, given by
\begin{equation}
    L_1=\alpha_1 \mathcal{L}_{rec}+(1-\alpha_1)\mathcal{L}_{sem},
\end{equation}
where $\alpha_1 \in [0,1]$ is a hyper-parameter that controls the trade-off between the two part loss functions. For instance, we can set to $\alpha=k/n$ in the practical semantic communication system. In this context, the system prioritizes the preservation of semantic information over the reconstructed quality when the bandwidth compression is small. In contrast, as the bandwidth compression increases, the system shifts its focus towards maintaining the reconstructed quality.

In the second training stage, we aim to further optimize the performance of the semantic communication system by jointly fine-tuning the encoder, decoder, and classifier with a small learning rate to achieve considerable inference performance and reconstructed image quality. One reason of fine-tuning the classifier is that the weights of the backbone and classifier are typically trained without considering channel corruption, which \replaced{causes}{leads} that the outputs of the backbone network may undergo substantial changes when \added{the} reconstructed images are inputted instead of the original images. This can result in a performance degradation. Therefore, fine-tuning the classifier with the semantic encoder and decoder can mitigate this issue and help enhance the semantics transmission. The loss function of this stage can be expressed as
\begin{equation}
    L_2=\alpha_2 \mathcal{L}_{rec}+(1-\alpha_2)\mathcal{L}_{Task},
\end{equation}
where $\alpha_2 \added{\in [0,1]}$ is a hyper-parameter like $\alpha_1$ and $\mathcal{L}_{Task}$ is the loss function of the downstream task. Specifically, when the downstream task is a classification problem, the cross-entropy function can be employed to model the loss, given by\deleted{:} 
\begin{equation}
        \mathcal{L}_{Task} = \mathbb{E}_{ \bm{x} \in \mathcal{B}} \bigg \{-\frac{1}{N_{cls}}\sum_{i=1}^{N_{cls}} y_{i} \log(\hat{y}_{i}) \bigg \},
\end{equation}
where $y_i$ and $\hat{y}_i$ represent the ground-truth and the predicted probability of the $i$-th class, respectively. Notation $N_{cls}$ denotes the number of classes in the dataset. 

% Motivated by the fact of transfer learning and  

\section{Simulations}
\begin{figure}[t!]
         \centering
         \begin{subfigure}{0.3\textwidth}
         \includegraphics[width=\textwidth]{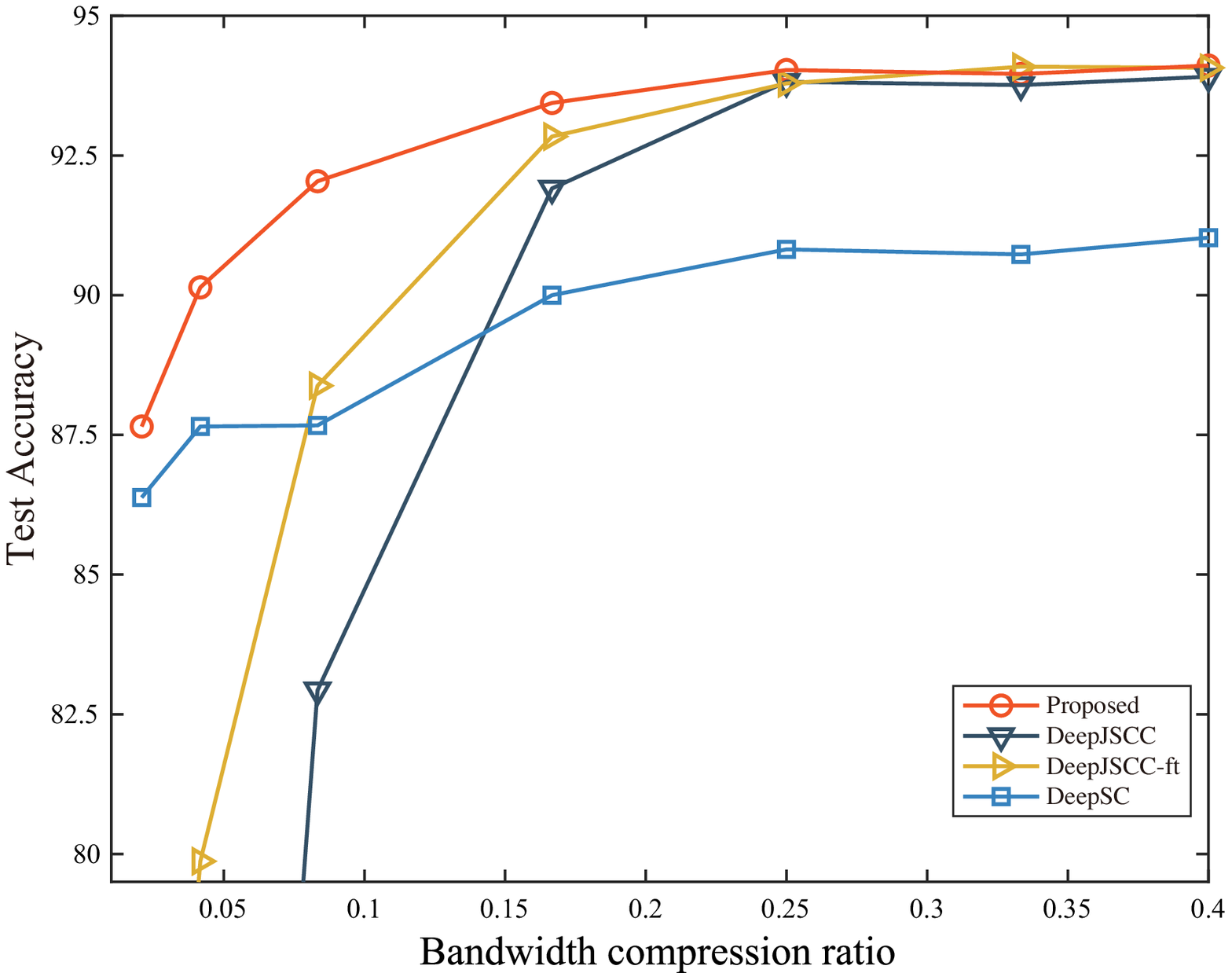}
         \caption{}
         \label{fig:cifar_snr20_acc}
     \end{subfigure}
     \begin{subfigure}{0.3\textwidth}
         \centering
         \includegraphics[width=\textwidth]{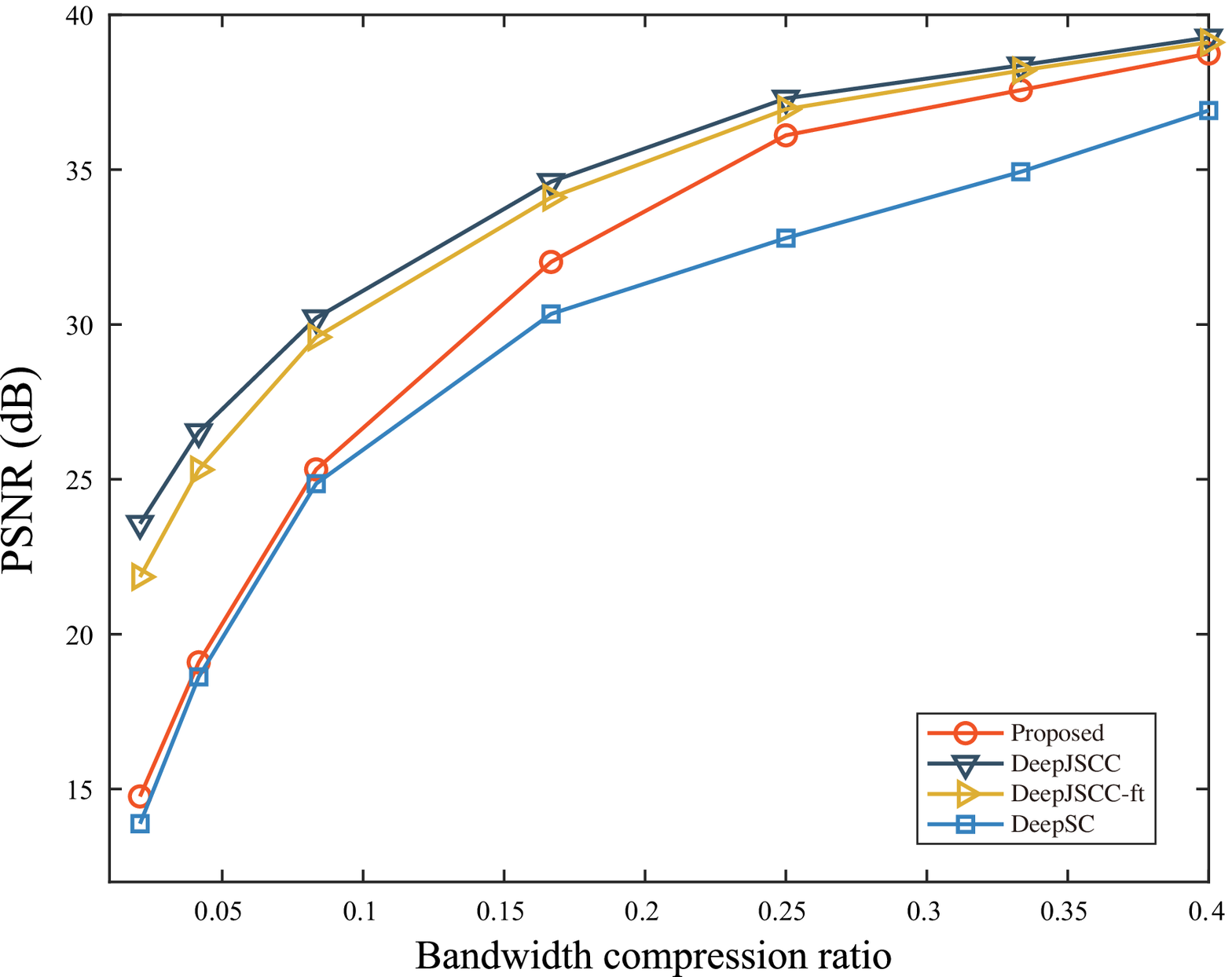}
         \caption{}
         \label{fig:cifar_snr20_psnr}
     \end{subfigure}
    \caption{Performance on CIFAR-10 versus the bandwidth compression ratio \replaced{with}{where} \deleted{the} SNR \replaced{=}{is} 20dB.}
        %\label{fig:three graphs}
\end{figure}
In order to verify the effectiveness of the proposed framework, we conduct experiments on CIFAR-10, which contains 60,000 32$\times$32 color images divided into 10 classes. The training set comprises 50,000 images, while the test set contains 10,000 images. A pre-trained ResNet-56 \cite{ResNet} \replaced{is}{was} used as the backbone network and classifier for downstream inference, while the structure of the semantic encoder and decoder is shown in Fig. \ref{fig:network}. The projection network adopts a two-layer fully connected structure with an output dimension of 32. The number of training epochs for the two stages \replaced{is}{was} set to 200 and 100, respectively, with a batch size of 128. Besides, we use the Adam optimizer with a learning rate of 0.01 for the first pre-training stage and 0.0001 for the second fine-tuning stage. These learning rates will be adjusted every 50 epochs with a decay factor of 0.5.
% \begin{figure*}[t!]
% 	\centering
% 	\subfloat[Test accuracy versus the bandwidth compression ratio where SNR is 20dB.]{\includegraphics[width=.25\textwidth]{figs/cifar_snr20.eps}}
% 	\subfloat[PSNR versus the bandwidth compression ratio where SNR is 20dB.]{\includegraphics[width=.25\textwidth]{figs/cifar_snr20_psnr.eps}}
% 	\subfloat[Title]{\includegraphics[width=.25\textwidth]{figs/cifar_snr10.eps}}
% 	\subfloat[Title]{\includegraphics[width=.25\textwidth]{figs/cifar_snr10_psnr.eps}}
% 	\caption{Description.}
% \end{figure*}
We \replaced{compare}{compared} the proposed \replaced{method}{framework} with the following DL-based semantic communication methods,
\begin{itemize}
    \item DeepJSCC \cite{DeepJSCC}: DL-based source-channel joint coding that maps the original input to the channel input through the structure of an autoencoder.
    \item DeepJSCC-ft: an extension of DeepJSCC with the second stage fine-tuning strategy of our proposed method, in which the encoder, decoder and classifier are updated with a small learning rate. 
    \item DeepSC \cite{DeepSC}: a DL-based semantic coding framework that  trains the semantic encoder and decoder with both semantic and observation losses to achieve efficient semantic information transmission. For a fair comparison, we freeze its encoder and decoder after training and retrained the classifier with a small learning rate.
\end{itemize}
% \begin{figure}[t!]
%     \centering
%     \includegraphics[width=2.8in]{figs/cifar_snr20.eps}
%     \caption{Test accuracy on CIFAR-10 testset versus the bandwidth compression ratio where the SNR is set 20dB.}
%     \label{fig:cifar_snr20_acc}
% \end{figure}
% \begin{figure}[t!]
%     \centering
%     \includegraphics[width=2.8in]{figs/cifar_snr20_psnr.eps}
%     \caption{PSNR on CIFAR-10 testset versus the bandwidth compression ratio where the SNR is set 20dB.}
%     \label{fig:cifar_snr20_psnr}
% \end{figure}
% \begin{figure}[t!]
%     \centering
%     \includegraphics[width=2.8in]{figs/cifar_snr10.eps}
%     \caption{Test accuracy on CIFAR-10 testset versus the bandwidth compression ratio where the SNR is set 10dB.}
%     \label{fig:cifar_snr10}
% \end{figure}
% \begin{figure}[t!]
%     \centering
%     \includegraphics[width=2.8in]{figs/cifar_snr10_psnr.eps}
%     \caption{PSNR on CIFAR-10 testset versus the bandwidth compression ratio where the SNR is set 10dB.}
%     \label{fig:cifar_snr10_psnr}
% \end{figure}

\begin{figure}[t!]
         \centering
         \begin{subfigure}{0.3\textwidth}
         \includegraphics[width=\textwidth]{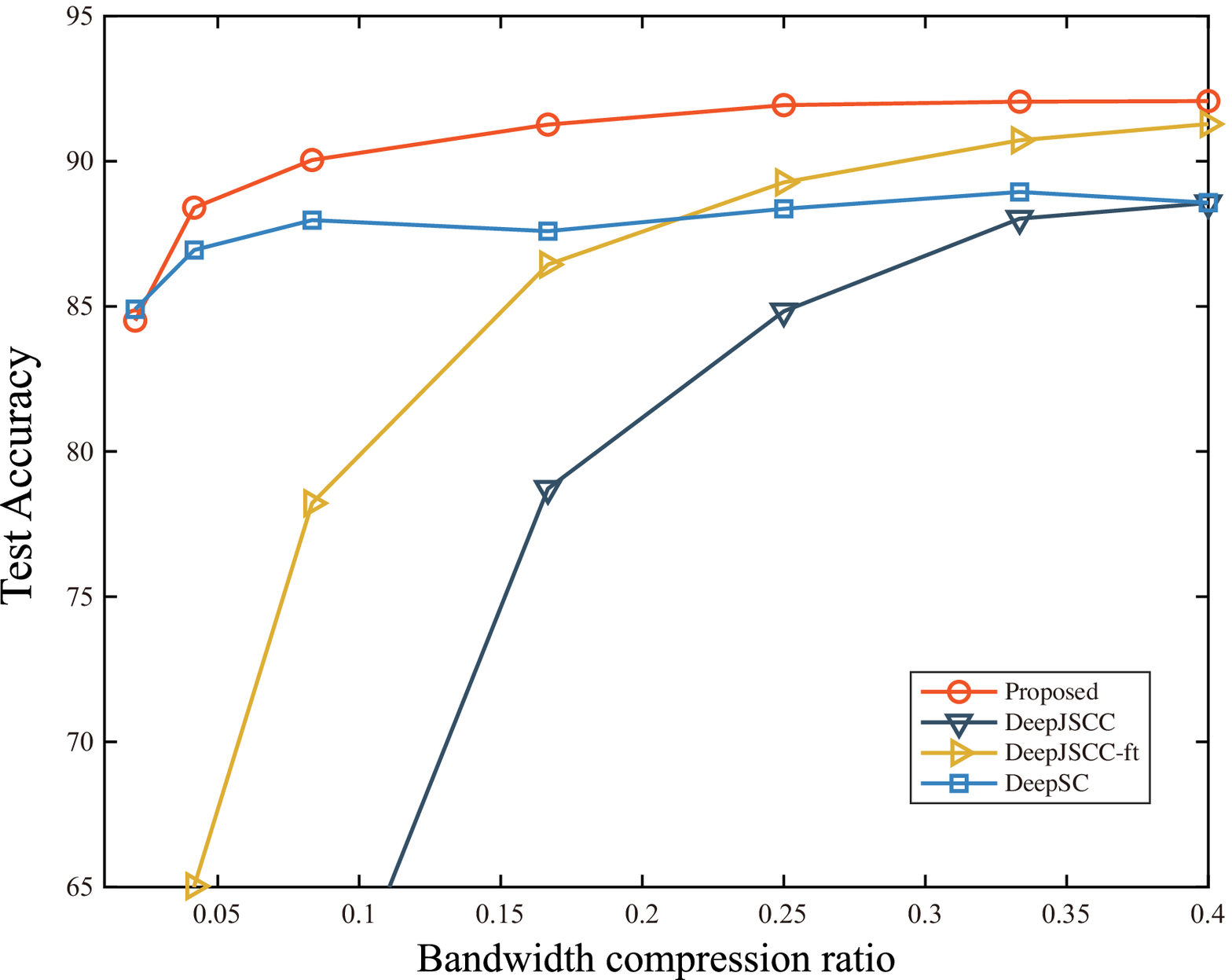}
         \caption{}
         \label{fig:cifar_snr5_acc}
     \end{subfigure}
     \begin{subfigure}{0.3\textwidth}
         \centering
         \includegraphics[width=\textwidth]{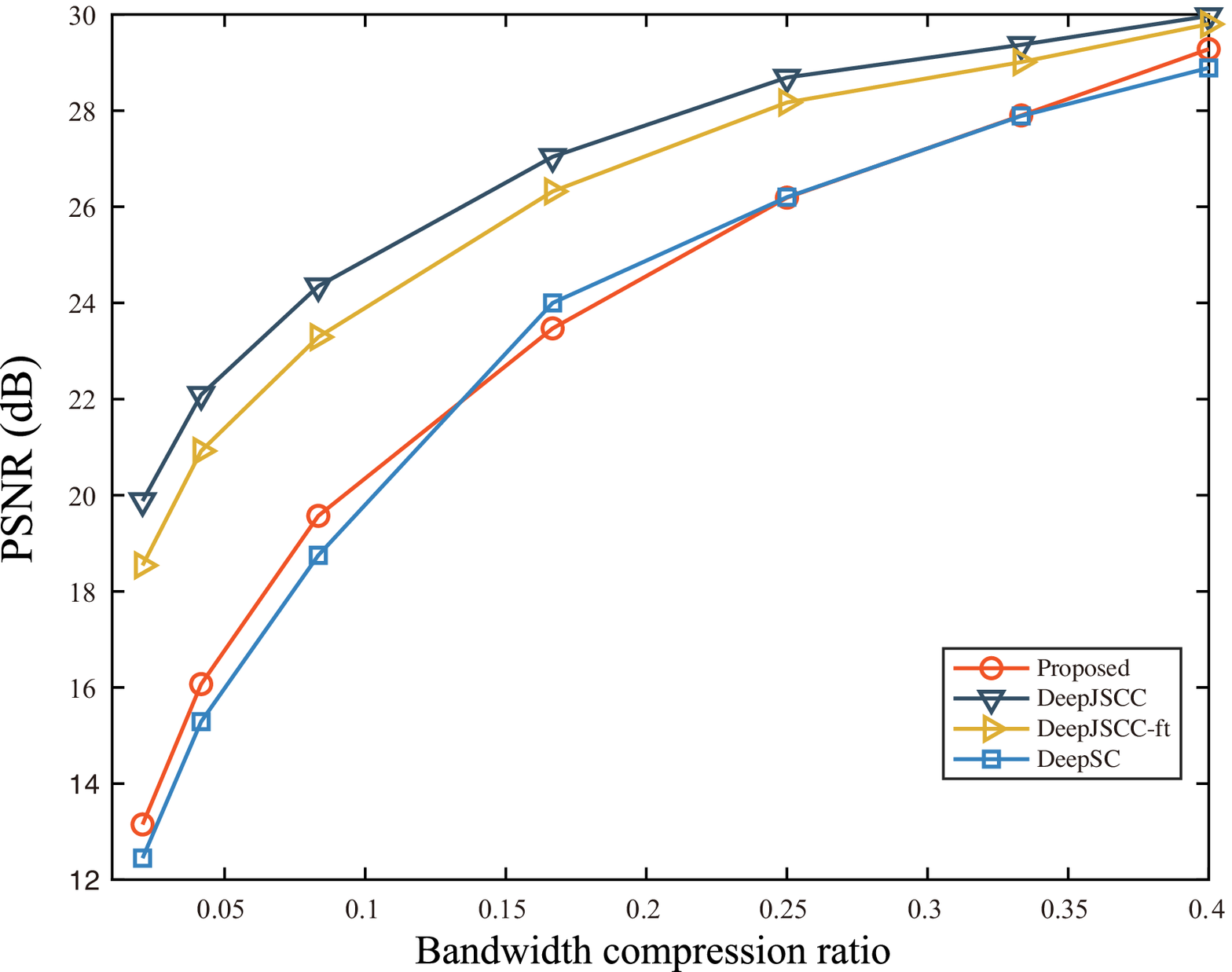}
         \caption{}
         \label{fig:cifar_snr5_psnr}
     \end{subfigure}
    \caption{Performance \deleted{comparison} on CIFAR-10 \deleted{testset} versus the bandwidth compression ratio \replaced{with}{where} \deleted{the} SNR \replaced{=}{is} 5dB.}
        %\label{fig:three graphs}
\end{figure}
\begin{figure*}[t!]
    \centering
    \includegraphics[width=0.88\textwidth]{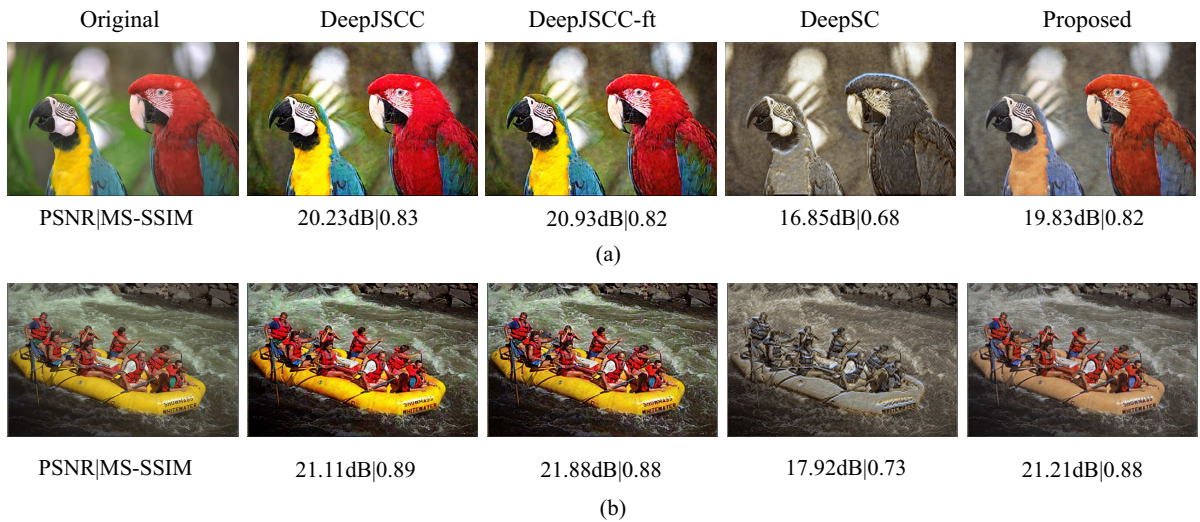}
    \caption{Visualization  comparison of \replaced{several}{different} methods on the Kodak dataset, where SNR \replaced{$=$}{is set to} 20dB and bandwidth compression ratio is 1/48.}
    \label{fig:vis}
    \vspace{-5mm}
\end{figure*}

 \autoref{fig:cifar_snr20_acc}  compares the accuracy performance of the proposed \replaced{method}{framework}, DeepJSCC, DeepJSCC-ft, and DeepSC, where SNR is set to 20dB, and the bandwidth compression ratio $k/n$ varies from 1/48 to 1/2.5. From this figure, we can find that the proposed method consistently outperforms or matches the \replaced[id=A2]{competitive}{compared} methods in accuracy. Specifically, when the compression ratio is 1/2.5, the proposed method, DeepJSCC, and DeepJSCC-ft achieve \replaced{the}{a} test accuracy of about 94\%, whereas DeepSC performs poorly with \replaced{the}{an} accuracy of only 91\%. As the bandwidth compression ratio decreases, \replaced[id=A2]{the proposed method still maintains a comparable accuracy performance. For instance, the proposed method can achieve accuracy of 90.14\% and 87.65\% at bandwidth compression ratios of 1/24 and 1/48, respectively, which outperforms DeepSC by about 2.5\% and 1\% at the corresponding bandwidth compression ratios and also shows an accuracy gain of up to 56\% over DeepJSCC.}{the accuracy performance of DeepJSCC declines dramatically, while DeepJSCC-ft performs better than DeepJSCC due to the use of fine-tuning. Moreover, DeepSC outperforms DeepJSCC and DeepJSCC-ft when the bandwidth compression ratio is extremely low.  Notably, the proposed method achieves accuracy of 90.14\% and 87.65\% at bandwidth compression ratios of 1/24 and 1/48, respectively, which outperforms DeepSC by about 2.5\% and 1\% at the corresponding bandwidth compression ratios.} These results indicate that the proposed framework can effectively extract semantic information to meet the requirements of downstream task and removes irrelevant redundant information to ensure the semantic information can be successfully transmitted, especially when channel bandwidth is limited.
 
  \autoref{fig:cifar_snr20_psnr} present\added{s} \replaced{the peak signal-to-ration (PSNR)}{the performance comparison} of the proposed and the three completing methods \deleted{in terms of PSNR}, where SNR is set to 20dB and the bandwidth compression ratio \deleted{$k/n$} varies from 1/48 to 1/2.5. As shown in the figure, we can find that as the bandwidth compression ratio increases, the PSNRs of all methods get improved. Although the proposed method \deleted{and DeepSC} sacrifice\added{s} some image quality to prioritize semantic information when the bandwidth compression ratio is low, \replaced{it}{they} can quickly catch up with DeepJSCC's PSNR at higher compression ratios. Specifically, the proposed \replaced{method}{framework} achieves a PSNR of 38.75dB, which is close to the 39.27dB of DeepJSCC, and outperforms DeepSC with 36.91dB when \added{the} bandwidth compression ratio is 1/2.5. These results indicate that the proposed \replaced{method}{framework} can prioritize to transmit semantic information over irrelevant background information to ensure the performance of downstream task in bandwidth-limited scenarios, \replaced{and meanwhile}{while} \replaced{transmit}{transmitting} enough background information to obtain good image quality when \added{the} bandwidth is not a bottleneck\replaced{.}{,} \replaced{This}{which} further demonstrates the effectiveness of the proposed \replaced{method}{framework}.
 
 \autoref{fig:cifar_snr5_acc} and  \autoref{fig:cifar_snr5_psnr} present the performance comparison of several methods under \deleted{a} poor channel condition\added{s} in terms of accuracy and PSNR, respectively. Specifically, both figures consider a low SNR of 5dB, and the bandwidth compression ratio \deleted{$k/n$} varies from 1/48 to 1/2.5. From \autoref{fig:cifar_snr5_acc}, we can observe that the proposed \replaced{methods}{framework} still shows the superiority in terms of accuracy compared to the three competitive methods, indicating its robustness in low SNR scenarios. From \autoref{fig:cifar_snr5_psnr}, we can find the proposed framework can adaptively sacrifice the global information to obtain comparable semantic performance when \added{the} bandwidth compression ratio is low, \replaced{and meanwhile}{while} obtain enough reconstructed quality in terms of PSNR as bandwidth compression ratio \replaced{decreased}{increased}. These results in both figures further verify the effectiveness and robustness of the proposed \replaced{method}{framework} in low SNR scenarios.
 
 We also provide the visualization  comparison of \replaced{several}{different} methods on the Kodak dataset in \autoref{fig:vis}, where the encoder and decoder are trained on the STL10 dataset, SNR is set to 20dB and the bandwidth compression ratio is 1/48. From this figure, we can observe that \replaced[id=A2]{the proposed can effectively  remove redundant background information and meanwhile preserve the main semantic information, resulting in less image corruption in semantic regions (e.g., macaws and rafters in this figures) compared to the competitive methods.  Moreover, the proposed method achieves similar PSNR and multi-scale structural similarity (MS-SSIM) performance with DeepJSCC and DeepJSCC-ft, indicating the effectiveness of the proposed method in reconstructing semantic information. These results further demonstrate the superiority of the proposed method in achieving leading accuracy in downstream tasks over the compared methods.
 }{DeepJSCC and DeepJSCC-ft do not focus on the main semantic information of the samples during transmission but treated each pixel equally, resulting in a significant consumption of bandwidth resources when transmitting complex background information, which caused distortion in the main semantic information of the reconstructed samples. Although DeepSC can effectively remove the redundant background information,  it can not sufficiently preserve the main semantic information in the reconstructed samples. However, the proposed method can remove redundant background information while preserving the main semantic information. By highlighting the main semantic information, the proposed method achieves similar MS-SSIM compared to those of DeepJSCC and DeepJSCC-ft. These phenomena can fully demonstrate the effectiveness of the proposed method in achieving leading accuracy in downstream task.}

\section{Conclusion}
In this paper, we proposed a CL-based semantic communication framework for wireless image transmission. The framework \replaced{incorporated}{incorporates} semantic contrastive coding and a two-stage training procedure to enhance the extraction of semantic information\added{,}  and \replaced{in order to}{and}achieve a better trade-off between reconstruction quality and the performance of downstream task. We evaluated the effectiveness of the proposed \replaced{methods}{framework} through simulations on CIFAR-10 and Kodak datasets, \replaced{which}{The} simulated results \replaced{demonstrated}{demonstrate} the superiority of our approach over existing methods.

\bibliographystyle{IEEEtran}

\bibliography{references}

%\newpage


%\vfill


\end{document}